\documentstyle[preprint,rotating,floats,epsf,aps]{revtex}

\begin{document}
\tighten
\def\beq{\begin{equation}}
\def\eeq{\end{equation}}
\def\beqy{\begin{eqnarray}}
\def\eeqy{\end{eqnarray}}
\def\pr#1#2#3{ {\sl Phys. Rev.\/} {\bf#1}, #3 (#2)}
\def\prl#1#2#3{ {\sl Phys. Rev. Lett.\/} {\bf#1}, #3 (#2)}
\def\np#1#2#3{ {\sl Nucl. Phys.\/} {\bf#1}, #3 (#2)}
\def\cmp#1#2#3{ {\sl Comm. Math. Phys.\/} {\bf#1}, #3 (#2)}
\def\pl#1#2#3{ {\sl Phys. Lett.\/} {\bf#1}, #3 (#2)}
\def\apj#1#2#3{ {\sl Ap. J.\/} {\bf#1}, #3 (#2)}
\def\aop#1#2#3{ {\sl Ann. Phy.\/} {\bf#1}, #3 (#2)}
\def\nc#1#2#3{ {\sl Nuovo Cimento }{\bf#1}, #3 (#2)}
\def\cjp#1#2#3{ {\sl Can. J. Phys. }{\bf#1}, #3 (#2)}
\def\zp#1#2#3{ {\sl Z. Phys. }{\bf#1}, #3 (#2)}
\def\pbi{\vec{p}_i}
\def\pbj{\vec{p}_j}
\def\yc{{\cal Y}}

\begin{titlepage}
\begin{flushright}
JLAB-TH-97-30
\end{flushright}
\vspace{1.cm}
\begin{center}
\Large\bf Meson Decays In A Quark Model\end{center}
\vspace{0.5cm}
\begin{center}
W. Roberts$^a$ and B. Silvestre-Brac$^b$ 
\\\vspace{3mm}
{\it  $^a$ Department of Physics, Old Dominion University\\
Norfolk, VA 23529 USA \\ and \\
Thomas Jefferson National Accelerator Facility\\
12000 Jefferson Avenue, Newport News, VA 23606, USA}\\\vspace{3mm}
{\it $^b$ Institut des Sciences Nucl\'eaires,\\
53 avenue des Martyrs, Grenoble, France}
\end{center}

\vspace{1.5cm}
\begin{abstract}
A recent model of hadron states is extended to include meson decays. We find
that the overall success of the model is quite good. 
Possible improvements to the model are suggested.
\end{abstract}
\end{titlepage}

\section{Introduction and Motivation}

Recently, a model for the description of hadronic states has been developed and
applied to both meson and baryon spectra \cite{bsb1,bsb2}. The success of the model in
describing the spectra in these sectors has led to attempts to take the model further by applying
it to multiquark states \cite{bsb3}. This success has also led to the question
of whether the model describes other aspects of baryon and meson spectroscopy
adequately. In particular, it is useful to know whether the model can help shed
light on some of the many outstanding puzzles in hadron spectroscopy. Before
such a question can be addressed, the model should be able to reproduce those portions of
hadron phenomenology that are better understood. In this article, we present 
the results of application of the wave functions obtained from this model to the
strong decays of light mesons. We find that the model is as successful as other
models, for the decays examined.

The rest of this article is organized as follows. In the next section we
summarize the salient points of the hadronic model, as well as the model we 
use to describe the decays of hadrons. In section 3 we
present our results with some discussion, while our conclusions are presented
in section 4.

\section{The Model}

Two ingredients are essential for a good description of meson decays. One of
these is a realistic description of the mesons themselves. The second is a
valid description of the transition operator. Both of these depend very much on
our educated guesses about non-perturbative QCD dynamics. In the case of the 
former, there
are many models proposed. We confine ourselves to the work of \cite{bsb1}. For
the meson decays, there are also a number of models proposed, but we will limit
ourselves to the pair creation model first proposed by Micu \cite{micu}, and popularized by
the Orsay group \cite{orsay}.

\subsection{Spectrum}

One possibility for the analysis of the meson spectrum is the use of a
`good' quark-antiquark potential. This is the course we pursue. Although the 
one-gluon exchange mechanism suggests possible functional forms for 
some terms of the
potential, the fact that we still do not know much about the mechanism of
confinement means that we must rely on phenomenology to determine both the form
and the parameters of the potential. Several potentials have been proposed in the past
in the literature. We believe that in models of the kind that we propose, a very crucial
test is a unified description of the meson and baryon spectra. This has been undertaken by
one of us \cite{bsb1,bsb2}, and six different potentials have been found which fulfill
this criterion. These potentials have been applied, not only to the meson and baryon
sectors, but also to tetra-quark states \cite{bsb3}, as well as to a description of the $KN$
interaction \cite{bsb4}. The results in all of these cases have been encouraging. 

In this paper, we select two of these potentials for this study. The general form of each
potential has been more or less imposed by
some basic QCD constraints, but the parameters have been determined by a fit
to a well-chosen sample of meson and baryon states. Both potentials are used with
a non-relativistic kinetic energy term, and they take the general form
\begin{equation}
V_{ij}(r)=-\frac{\kappa }{r}+\lambda r^{p}+\Lambda +\frac{2\pi \kappa
^{\prime }}{3m_{i}m_{j}}\frac{\exp (-r^{2}/r_{0}^{2})}{\pi ^{3/2}r_{0}^{3}}
\overrightarrow{\sigma _{i}}\overrightarrow{\sigma _{j}}.
\end{equation}

One peculiarity of these potentials is that the range $r_{0}$ of the
hyperfine term  is mass dependent through the
relation
\begin{equation}
r_{0}(m_{i}m_{j})\ =\ A\left( \frac{2m_{i}m_{j}}{m_{i}+m_{j}}\right) ^{-B}.
\end{equation}
In the limit of a vanishing range, this term gives the usual
Fermi-Breit prescription $\delta (\vec{r})$.

These potentials contain neither spin-orbit nor tensor contributions so that
$^3L_J$ states have the same wave functions. In the same spirit,
$^3L_J\leftrightarrow ^3L^\prime_J$ and $^3L_J\leftrightarrow ^1L_J$ mixings
are absent. This approximation can have some effect on the results, but we are
mainly interested in the gross features of the decays, so that the potentials
are quite satisfactory from this point of view.
Furthermore, there is no isospin dependence (this is a property shared by most
potential models: in \cite{GI}, isospin dependence is introduced by means of mixing through annihilation). Thus, the $\rho$ and $\omega$, as well as the $a_J$ and
$f_J$ resonances are degenerate, both in energy and in wave function. More
dramatic is the case of the pseudoscalar mesons, as this leads to a degeneracy
of the $\pi$ and the non-strange part of the $\eta$. Isospin dependence appears
only through the flavor part of the wave function, although there probably
should be a
difference in the spatial part of the wave function as well. In any case, we
consider two potentials that yield good overall results for the meson spectrum.

The first potential, denoted AL1, has the usual Coulomb+linear form for 
the central part, and its parameters are
\begin{eqnarray*}
m_{u} &=&m_{d}=0.315 {\rm GeV};\ m_{s}=0.577 {\rm GeV};\ m_{c}=1.836 {\rm GeV};\
m_{b}=5.227 {\rm GeV} \\
\kappa  &=&0.5069;\ \kappa ^{\prime }=1.8609;\ \lambda =0.1653 {\rm GeV}^{2};
\ p=1;\ \Lambda =-0.8321 {\rm GeV} \\
B &=&0.2204;\ A=1.6553 {\rm GeV}^{B-1}.
\end{eqnarray*}

The second choice of potential, denoted AP1, has a confining term suggested by the
Regge trajectory behavior of orbital states in a non-relativistic treatment.
The parameters are
\begin{eqnarray*}
m_{u} &=&m_{d}=0.277{\rm GeV};\ m_{s}=0.553{\rm GeV};\ m_{c}=1.819{\rm GeV};\
m_{b}=5.206{\rm GeV} \\
\kappa  &=&0.4242;\ \kappa ^{\prime }=1.8025;\ \lambda =0.3898{\rm GeV}^{5/3};
\ p=2/3;\ \Lambda =-1.1313{\rm GeV} \\
B &=&0.3263;\ A=1.5296{\rm GeV}^{B-1}.
\end{eqnarray*}

Both potentials reproduce spectra of comparable quality and are simple enough to be
handled without any difficulty. In particular, it is quite easy to solve the
differential equation resulting from the Schr\"{o}dinger equation. However,
the radial part $R_{nls}(r)$ of the meson wave function is calculated numerically on a
grid. Such a form is not easily used for studying more complicated problems: a 
continuous expression is usually more convenient.

\begin{table}[tbp] \centering
\begin{tabular}{cccccc}
system &\multicolumn{5}{c}{masses (MeV)} \\ \cline{2-6}
 & experiment & exact & $N=1$ & $N=2$ & $N=3$ \\ \hline
$\pi $ & 138 & 138 & 194 & 140 & 138 \\ 
$\pi (1300)$ & 1300 & 1303 & - & 1314 & 1303 \\ 
$\rho $ & 769 & 770 & 771 & 770 & 770 \\ 
$\omega $ & 782 & 770 & 771 & 770 & 770 \\ 
K & 496 & 490 & 532 & 492 & 491 \\ 
K$^{*}$ & 893 & 903 & 906 & 904 & 903 \\ 
$\phi $ & 1019 & 1020 & 1025 & 1021 & 1020 \\ 
a$_{J}$ & 1251 & 1208 & 1210 & 1208 & 1208 \\ 
f$_{J}$ & 1245 & 1208 & 1210 & 1208 & 1208
\end{tabular}
\caption{Masses of some mesons used in the paper obtained with potential
 AL1. The experimental values are given in the second column; for the
 $a_J$ and $f_J$ resonances we show the energy centroid. The third
 column reports the exact values obtained from the Schr\"odinger equation
 by solving the resulting differential equation. The last three columns
 deal with the approximation based on gaussian functions with, respectively,
 $N=1,2,3$ terms in the expansion.\label{table0}}
\end{table}

\begin{figure} 
\caption{The pion wave function that results from two choices of potential, with different
expansion approximations. The curves on the left are from potential AL1, while those on
the right are from AP1. \label{fig1}}
\centerline{\mbox{\begin{turn}{90}%
\epsfysize=9.0cm\epsffile{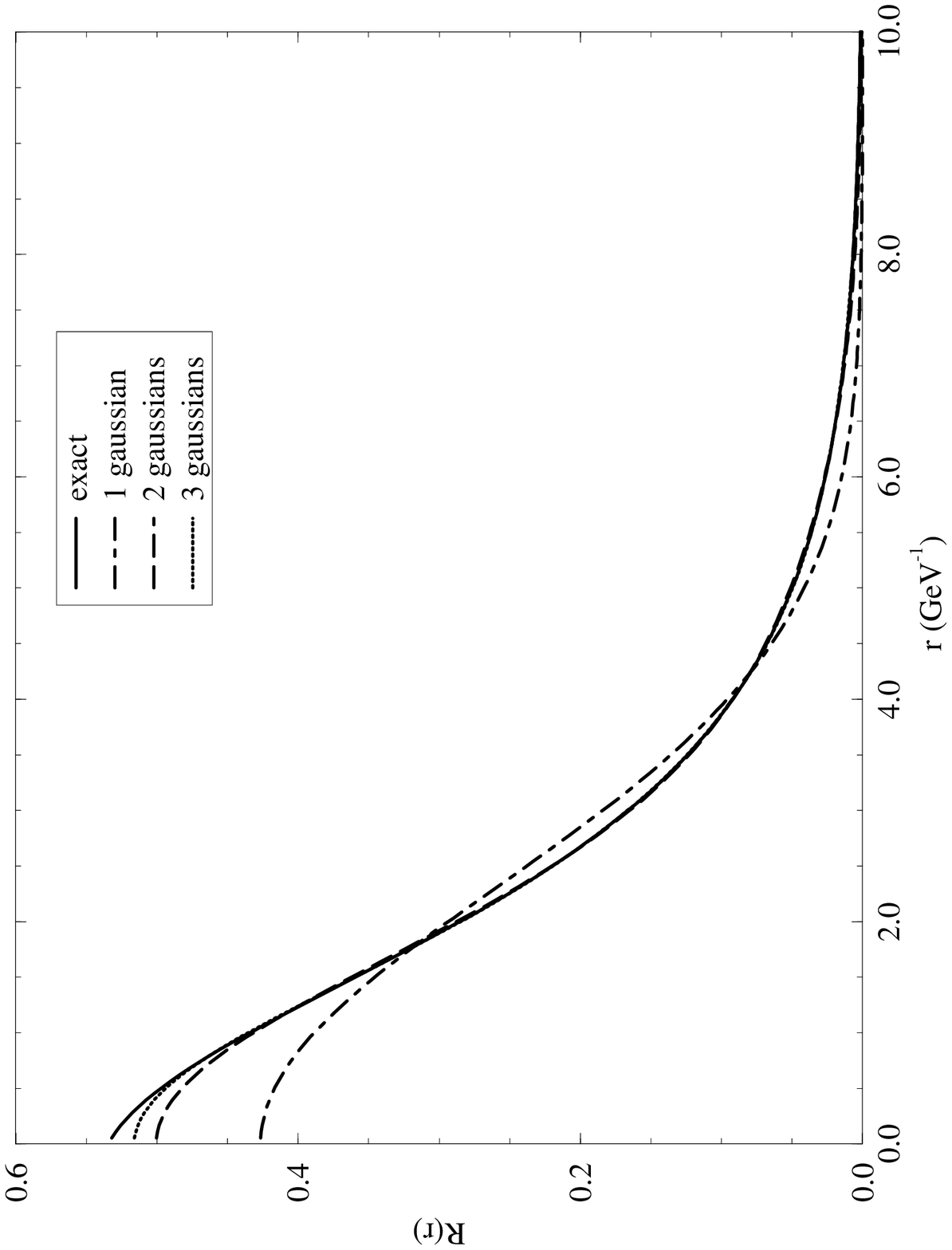}\end{turn}}\hskip 0.5cm\mbox{\begin{turn}{90}%
\epsfysize=9.0cm\epsffile{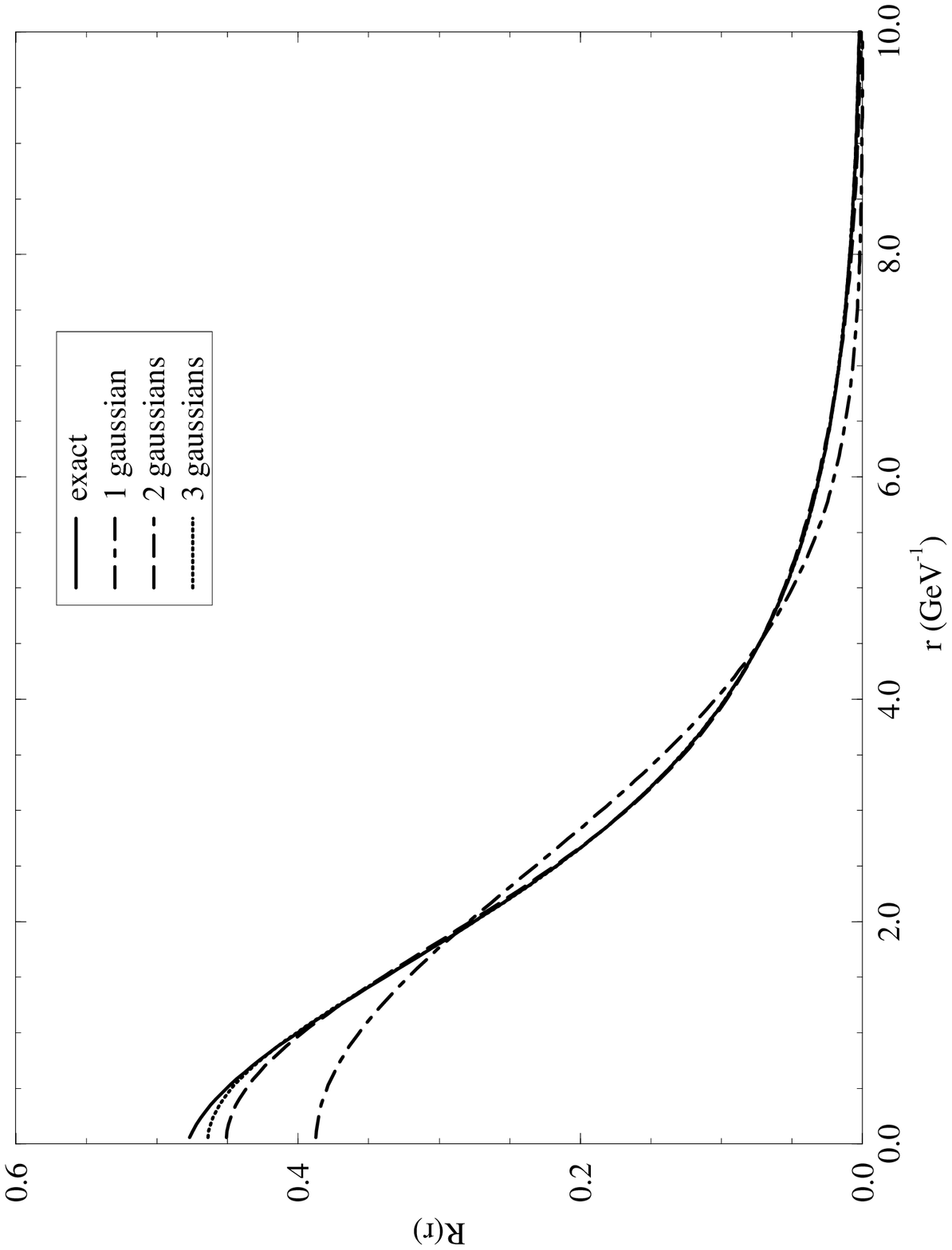}\end{turn}}}
\end{figure}

To solve this problem, we approximate the regularized part of
the exact radial wave function by a linear combination of gaussian functions,
\begin{equation}\label{expansion}
R_{nls}(r)=\sum_{i=1}^{N}c_{i}r^{l}\exp (-\alpha _{i}r^{2}/2).
\end{equation}
For a given number $N$ of gaussian functions, the parameters $c_{i}$ and 
$\alpha _{i}$ are determined by a variational procedure on the energy of the
considered state. $N=1$ is a rather rough approximation, but $N=2$ or $N=3$ greatly improves 
the results. This is illustrated in figure \ref{fig1}, where the exact wave function, as well as the 
$N=1,2,3$ approximations, are plotted for the pion, for a large range of $r$
values. In this figure, the curves on the left are for the potential AL1, while those on
the right are for AP1. We see that except for very short distances, the
approximation with 3 gaussian terms gives essentially the exact wave
function. This expression, eqn. (\ref{expansion}), is then used in our model for meson
decays.

We show in table \ref{table0} a few of the masses obtained in this way, for the potential Al1. Experimental values are
shown in the second column, while the exact predictions are given in the third. The masses obtained using the expansion
of the wave function in gaussian functions are shown in the next three columns for $N=1,2,3$, respectively.
We see that the approximation with $N=3$ gives essentially the same results as
the exact case, confirming the trend illustrated in figure \ref{fig1}.

\subsection{Decays}

For the OZI \cite{OZI} allowed decays of a meson, we assume that quark pair creation,
somewhere within the volume of the parent hadron, provides the dominant
contribution. In fact, for the operator we choose, this creation can take place
anywhere in space, but the wave function overlaps ensure that the dominant
contribution to the amplitude comes from creation within the `confines' of the
parent hadron.

The full mechanism for strong decays is, of course, as yet unknown, but the
success of pair creation models hints at what may be the most important
aspects of the underlying dynamics. We assume that the operator responsible for
the decay is
\beqy \label{amp1}
T&=&-3\gamma\sum_{i,j} \int d \pbi\/ d \pbj\/ \delta( \pbi\/ +
\pbj\/) C_{ij} F_{ij}\nonumber \\
&\times& \sum_m <1,m;1,-m|0,0> \chi_{ij}^m { \cal Y\/ }_1^{-m}( \pbi\/
- \pbj\/ )b_i^{\dagger}( \pbi\/ ) d_j^{\dagger}( \pbj\/ ).
\eeqy
Here, $C_{ij}$ and $F_{ij}$ are the color and flavor wave functions of the 
created pair, both assumed to be singlet, $\chi_{ij}$ is the spin triplet wave 
function of the pair, and ${\cal Y}_1(\pbi - \pbj)$ is the vector harmonic 
indicating that the pair is in a relative P-wave. $b_i^\dagger$ creates a quark, and
$d_i^\dagger$ creates an antiquark.

For the transition $A\to B C$, we are interested in evaluating the transition 
amplitude $M$, given by
\beq
M=<BC|T|A>.
\eeq
The details of this calculation are given elsewhere \cite{RB,scwr}. Once $M$ is evaluated, we
calculate the decay rate as 
\beq
\Gamma(A\to BC)=\sum_{J_{bc},\ell}\left|M_{A\to BC}(J_{bc},\ell,k_0)\right|^2\Phi(ABC),
\eeq
where $\Phi$ is the phase space for the decay. We have investigated three
possible choices for this phase space. The first choice is that of fully
relativistic phase space, for which 
\beq
\Phi(ABC) = 2\pi 
{E_b(k_0) E_c(k_0) k_0 \over m_a},
\eeq
with $E_i(k_0)=\sqrt{ k^2_0 + m_i^2}$, and $k_0$ is calculated in the rest frame of the parent
meson. The second choice is fully
non-relativistic phase space, with
\beq
\Phi(ABC) = 2\pi {m_b m_c k_0 \over m_a}.
\eeq
Our third choice is the `weak-binding' phase space suggested by Kokoski and
Isgur \cite{KI}, for which
\beq \label{pspace1}
\Phi(ABC) = 2\pi {\tilde m_b \tilde m_c k_0 \over \tilde m_a},
\eeq
where the $\tilde m$'s are effective meson masses (in the model) without
hyperfine interaction. For the $\tilde{m}_i$, we use the values suggested in
\cite{KI}.

\section{Results and Discussion}

We have performed a number of fits to the experimental data; we show the results of these fits in table
\ref{tablea}. In addition, using the parameters obtained from these fits, we have calculated the amplitudes
for a number of other decays. These are presented in tables \ref{tableb}, \ref{tablec} and \ref{tabled}. In table
\ref{tablea}, we consider only decays in which the daughter mesons are narrow. For this purpose, we treat the $\omega$ as a 
narrow state. In tables \ref{tableb} to \ref{tabled}, we extend the calculation to include $\rho$'s and $K^*$'s
as daughter mesons.

\squeezetable

\begin{table} %\label{tablea}
\caption{Results from various fits. Fit I corresponds to a fit to the single decay $\rho\to\pi\pi$, using the
simple form of the $^3P_0$ operator. Fit II corresponds to the same choice of operator, but to a global fit to
all of the decays in this table. Fit III also results from a global fit, using the modified form of the $^3P_0$
operator. Fits I, II and III result from single-component, AL1 wave functions. Fits IV to VII are all 
global fits to all the decays shown in the table. Fit IV results
from using the three-component, AL1 wave functions, with the usual form of the $^3P_0$ operator.
Fit V uses the same wave functions and the modified $^3P_0$ operator. Fit VI and VII also use this form of the operator, but
Fit VI uses the single component form of the AP1 wave functions, while Fit VII uses the
three-component form of the AP1 wave functions. All widths are calculated using the weak-binding phase space
prescription. The experimental results are from \protect{\cite{pdg}}. \label{tablea}}
\begin{center}
\begin{tabular}{||lrrrrrrrr||}\hline
\multicolumn{1}{||c}{decay}&\multicolumn{8}{c||}{$\Gamma^{1/2}$ (MeV$^{1/2}$)}\\ \cline{2-9}
 & Fit I & Fit II & Fit III & Fit IV & Fit V & Fit VI & Fit VII & experiment\\ \hline\hline
$\rho\to\pi\pi$         & 12.3 & 11.1 & $11.9\pm 0.3$ & 10.4 & 12.1 $\pm$ 0.3& 12.0 $\pm$ 0.3 &   12.2 $\pm$ 0.3  & $12.3\pm 0.5$ \\ \hline
$K^*\to K\pi$           & 8.6  & 7.8 & $7.8\pm 0.2$ & 7.2 & 7.5 $\pm$ 0.2& 7.7 $\pm$ 0.2&   7.5 $\pm$ 0.2  &  $7.1 \pm 0.4$ \\ \hline
$\phi\to K \bar{K}$     & 2.8 & 2.5 & $2.5\pm 0.1$ &  2.3 & 2.3 $\pm$ 0.1&2.4 $\pm$ 0.1  &  2.3 $\pm$ 0.1 &$1.9\pm 0.2$ \\ \hline
$f_2(1275)\to\pi\pi$    & 9.5  & 8.6 & $8.7\pm 0.3$ & 9.2 &  8.4 $\pm$ 0.3& 9.2 $\pm$ 0.3& 9.4 $\pm$ 0.2  &$12.5 \pm 1.3$ \\ \hline
$f_2(1275)\to K\bar{K}$ & 2.5 & 2.3 & $1.7\pm 0.1$ & 2.3 &1.4 $\pm$ 0.1  & 1.6 $\pm$ 0.1 &   1.4 $\pm$ 0.0 &$2.9\pm 0.4$ \\ \hline
$f_2(1275)\to\eta\eta$  & 0.7  & 0.8 & $0.4\pm 0.4$ & 0.7 &0.3 $\pm$ 0.0  & 0.4 $\pm$ 0.0&0.3 $\pm$ 0.0   &$0.9 \pm 0.2$ \\ \hline
$f_0(1300)\to\pi\pi$          & 11.5 & 10.5 & $15.5\pm 1.7$ & 5.5 & 19.6  $\pm$ 1.6& 17.2 $\pm$ 1.6&  16.5 $\pm$ 0.4      &$16.0\pm 6.7$ \\ \hline
$f_0(1300)\to K\bar{K}$       & 9.5  & 8.6  & $10.3\pm 0.4$ &6.3 & 9.6 $\pm$ 0.4& 9.7 $\pm$ 0.3 &  8.7 $\pm$ 0.2    & $4.5 \pm 2.1$ \\ \hline
$a_2(1320)\to\eta\pi$   & 2.0 & 1.8 & $1.7\pm 0.1$ &2.0 & 1.5 $\pm$ 0.1& 1.4 $\pm$ 0.1& 1.6 $\pm$ 0.0 & $3.9\pm 0.4$ \\ \hline
$a_2(1230)\to K\bar{K}$ & 4.2  & 3.8 & $3.0\pm 0.2$ &3.9 &2.6 $\pm$ 0.2  & 2.6 $\pm$ 0.1& 2.7 $\pm$ 0.1 &  $2.3 \pm 0.3$ \\ \hline
$f_2(1525)\to K\bar{K}$ & 9.4 & 8.5 & $8.4\pm 0.3$ & 8.1 &7.5 $\pm$ 0.2  & 8.3 $\pm$ 0.3& 8.1 $\pm$ 0.2 &$7.3\pm 1.6$ \\ \hline
$f_2(1525)\to\eta\eta$  & 2.6  & 2.4 & $2.3\pm 0.1$ &2.4 & 1.9 $\pm$ 0.1& 2.1 $\pm$ 0.1 &  2.1 $\pm$ 0.1 & $4.6 \pm 1.1$ \\ \hline
$K_1(1275)\to K\omega$  & 7.2 & 6.5 & $6.1\pm 0.2$ &5.5 & 5.4 $\pm$ 0.1 & 5.3 $\pm$ 0.1& 4.8 $\pm$ 0.1  & $3.1\pm 0.9$ \\ \hline
$K_1(1400)\to K\omega$  & 1.9  & 1.7 & $1.9\pm 0.1$ & 1.4 &1.7 $\pm$ 0.0 & 1.6 $\pm$ 0.0& 1.5 $\pm$ 0.0  &$1.3 \pm 0.8$ \\ \hline
$K^*(1410)\to K\pi$     & 1.3 & 1.2 & $1.1\pm 0.2$ & 0.1 &1.3 $\pm$ 0.1 & 2.6 $\pm$ 0.3 &  1.9 $\pm$ 0.0  &$3.9\pm 0.8$ \\ \hline
$K_0(1430)\to K\pi$     & 7.5  & 6.8 & $10.4\pm 1.1$ &  3.3 & 12.5 $\pm$ 1.0& 11.1 $\pm$ 1.0 &  10.3 $\pm$ 0.3  &$16.3 \pm 2.0$ \\ \hline
$K_2(1425)\to K\pi$     & 7.8 & 7.1 & $7.1\pm 0.2$ & 7.3 &6.8 $\pm$ 0.2 &  6.7 $\pm$ 0.2 &  7.4 $\pm$ 0.2  &$7.2\pm 1.5$ \\ \hline
$K_2(1425)\to K\omega$  & 2.1  & 1.9 & $1.3\pm 0.1$ & 1.8 &1.1 $\pm$ 0.1 & 1.2 $\pm$ 0.1&  1.1 $\pm$ 0.0 &$1.7 \pm 0.6$ \\ \hline
$K_2(1425)\to K\eta$    & 4.2 & 3.6 & $3.1\pm 0.2$ &4.0 & 2.7 $\pm$ 0.2 & 2.5 $\pm$ 0.2& 2.7 $\pm$ 0.1  & $0.4\pm 0.5$ \\ \hline
$K_1(1680)\to K\pi$     & 3.5  & 3.2 & $5.1\pm 0.4$ &  1.5 & 4.9 $\pm$ 0.4 &5.0 $\pm$ 0.3  &  3.9 $\pm$ 0.1  &$11.2 \pm 3.0$ \\ \hline
$K_3(1780)\to K\pi$     & 4.8 & 4.3 & $4.3\pm 0.2$ & 5.1 & 4.6 $\pm$ 0.2 & 3.7 $\pm$ 0.2&  5.3 $\pm$ 0.1   &$5.6\pm 0.8$ \\ \hline
$K_3(1780)\to K\eta$    & 3.0  & 2.7 & $2.1\pm 0.2$ &3.3 & 2.0 $\pm$ 0.2 & 1.4 $\pm$ 0.2&  2.2 $\pm$ 0.1    & $3.6 \pm 0.8$ \\ \hline
\end{tabular}
\end{center}
\end{table}

The numbers in column two (Fit I) of these tables result from a fit to the amplitude for $\rho\to\pi\pi$, using the
usual form of the $^3P_0$ operator. The numbers in column three (Fit II) result from using the same operator, but
fitting to the 22 decays shown in table \ref{tablea}. Column four (Fit III) corresponds to a similar global fit, but
using a modified transition operator. In this case, the operator has been modified by
making the change
\begin{eqnarray}
&&\gamma\sum_m\chi^m {\cal
Y}_1^{-m}\left(\vec{p}\right)<1,m;1,-m|0,0>\nonumber\\
&&\to
e^{-\frac{\lambda^2p^2}{2}}\left(\gamma_1+\gamma_2p^2\right)\sum_m\chi^m 
{\cal Y}_1^{-m}\left(\vec{p}\right)<1,m;1,-m|0,0>,
\end{eqnarray}
with $\lambda=1.37$, $\gamma_1=0.14$ and $\gamma_2=1.4$. For the results of
column 2 and 3, we find $\gamma=0.41$ and $\gamma=0.37$, respectively.  These three fits all result from the AL1
wavefunctions, expanded to $N=1$.

The effects of the size of the expansion basis of the wave
functions are illustrated in Fits IV and V; the former is a one parameter global fit
($\gamma=0.31$) using the
AL1 wave functions with three gaussians, while the latter uses the same wave functions with the
modified transition operator ($\lambda=0.0$, $\gamma_1=0.14$, $\gamma_2=0.94$). Fits VI and VII 
show the effects of a change in potential. Both of
these arise from the AP1 wave functions in conjunction with the modified transition operator: Fit
VI uses wave functions with a single gaussian ($\lambda=1.1$, $\gamma_1=0.13$, 
$\gamma_2=1.3$), while the wave functions of Fit VII are expanded
in a basis of three gaussians ($\lambda=0.0$, $\gamma_1=0.10$, $\gamma_2=1.0$). For Fits III, V,
VI and VII, the theoretical errors are evaluated using the covariance matrix resulting from
the fit. All of the amplitudes are calculated using the weak-binding prescription for phase space. As can be seen from tables \ref{tablea}, there
are few instances in which a change of the choice of potential or wave function expansion,
accompanied by apparently radical changes in fit parameters, induces a big change in the amplitude.
In particular, the modification of the operator induces an enhancement of large
decay amplitudes, while keeping the others at approximately constant values,
thus increasing the agreement with the experimental data. The most noticeably different fit is Fit IV.

The effective of modifying the wave function is less obvious, although the wave
functions expanded on a larger basis ($N=3$) gives slightly better fits than
those expanded on the smaller basis.

We have also fitted using only the first three decays of table \ref{tablea}, namely
$\rho\to\pi\pi$, $K^*\to K\pi$ and $\phi\to KK$. We have also omitted various ones of the decays in
table \ref{tablea} from the fit, such as the decays of the $f_0(1300)$. Except for a very few isolated amplitudes, 
none of these modifications have had a significant effect on the decay amplitudes that result. 
We note also that non-relativistic phase space consistently gives by far the worst results,
while the weak-binding prescription gives the best results. However, we find that in the
case of the full relativistic phase space, there are two decays that consistently
contribute the most to the `badness of fit'. One is the decay $\rho\to\pi\pi$, while the
other is $K_2(1425)\to K\eta$. But for these two decays, relativistic phase space and the
weak-binding prescription would give fits with essentially the same values of $\chi^2$.

The wave functions that we use take into account SU(3) breaking effects. Strictly speaking we
should also fully include such effects in the amplitudes. These effects arise from three sources.
The most obvious of these is in the calculation of phase space, where we use either the physical 
masses of the states (for non-relativistic and relativistic phase space), or the `weak-binding'
masses appropriate to the mesons in the decay. The second source of SU(3) breaking is due to the
fact that the amplitude depends on the momentum of the daughter mesons, calculated in the rest
frame of the parent. Here we also use the physical masses of the states to calculate this momentum.
The third source of SU(3) breaking arises in the evaluation of the matrix element of the transition 
operator. The masses of the quarks or, more precisely, various ratios of masses of the quarks,
enter explicitly into this calculation. We have found that treating the $u$, $d$ and $s$ as
degenerate provides the best results. Thus, SU(3) breaking at this level is not necessary.

A few comments on the results of table \ref{tablea} are in order. The amplitudes for the $K_1(1275)$ and
$K_1(1400)$ are calculated from those for the $^3P_1$ and $^1P_1$ states of the model spectrum, assuming a
mixing angle of $34^\circ$. In tables \ref{tableb} to \ref{tabled}, we ignore this mixing and present results
for the unmixed $^3P_1$ and $^1P_1$. The $\eta$ and $\eta^\prime$ are assumed to be
\begin{equation}
\eta=\frac{1}{\sqrt{2}}\left[\frac{1}{\sqrt{2}}\left(u\bar{u}+d\bar{d}\right)-s\bar{s}\right],
\,\,\, \eta^\prime=\frac{1}{\sqrt{2}}\left[\frac{1}{\sqrt{2}}\left(u\bar{u}+d\bar{d}\right)+s\bar{s}\right],
\end{equation}
respectively. The $\omega$ is assumed to be purely $u\bar{u}+d\bar{d}$, and the $\phi$ is taken as pure
$s\bar{s}$. Finally, in all of the results we have ignored mixings of the type
$^3L_J\leftrightarrow ^3L^\prime_J$, while in tables \ref{tableb} to \ref{tabled}, we have also ignored mixings of
the type $^1L_J\leftrightarrow ^3L_J$.

The numbers in table \ref{tablea} show that the model reproduces the experimental data reasonably well, for
all of the scenarios presented. Not surprisingly, the three-parameter fits are somewhat better than any of
the one-parameter fits. With few exceptions, amplitudes that are found to be large experimentally, also turn
out to be large in the model, and the same is generally true of small amplitudes. The few exceptions where a
large discrepancy exists are $f_0\to KK$, $K_1(1275)\to K\omega$ and $K_1(1680)\to K\pi$. In the case of the
first of these decays, we note simply that the status of this state is very much in question, so that the
partial widths themselves are also uncertain. In the case of the $K_1(1275)$, we note that the mixing we have
used has not resulted from a `rigorous' calculation, but is simply `borrowed' from the literature. In the
case of the $K_1(1680)$, there is the possibility of mixing with the radially excited $^3S_1$ state, not
taken into account in this calculation. However, indications are that this mixing is very small \cite{GI}, and would
probably not play a large role in bringing the model calculation into closer agreement with the experimental
measurement.

\begin{table} 
\caption{Results from various fits, for states with $I=1$. The key is as in table \protect{\ref{tablea}}. 
\label{tableb}}
\begin{center}
\begin{tabular}{||rlrrrr||}\hline
\multicolumn{2}{||c}{decay}&\multicolumn{4}{c||}{$\Gamma^{1/2}$ (MeV$^{1/2}$)} \\\cline{3-6}
& & Fit I & Fit II & Fit III & experiment\\ \hline\hline
$a_1(1235)\to$ & $ (\rho\pi)_{L= 0}$ &  9.3$\pm$  1.0 &  8.4$\pm$   0.9 &  9.1$\pm$  1.2 & dominant 
\tablenote{Experimental values are for the sum over all relevant partial waves.}\\
 & $ (\rho\pi)_{L= 2}$ &   3.1$\pm$   0.7 &   2.8$\pm$   0.6 &   2.1$\pm$   0.5 & \\
 & $KK^*$ &   0.3$\pm$   0.3 &   0.3$\pm$   0.2 &   0.2$\pm$   0.2 & possibly seen\\ \hline
$a_2(1320)\to$ & $KK$ &   4.2$\pm$   0.2 &   3.8$\pm$   0.1 &   3.0$\pm$   0.1 & $2.3 \pm 0.3$ \\
 & $\eta\pi$ & 2.0$\pm$   0.1 &  1.8$\pm$   0.1 &  1.7$\pm$   0.1 & $3.9\pm 0.4$  \\
 & $\eta^\prime\pi$ & 0.5$\pm$   0.1 &   0.4$\pm$   0.0 &   0.3$\pm$   0.0 & 0.9 $\pm$   0.1 \\
 & $\rho\pi$ &  5.2$\pm$   0.1 &  4.7$\pm$   0.1 &  3.7$\pm$   0.1 & 10.5 $\pm$   0.4\\
 & $(\rho\omega)_{S= 2,L= 0}$ &   0.6$\pm$   0.0 &   0.5$\pm$   0.0 &   0.4$\pm$   0.0 & --- \\
 & $(\rho\omega)_{S= 2,L= 2}$ &   0.1$\pm$   0.0 &   0.1$\pm$   0.0 &   0.1$\pm$   0.0 & --- \\
 & $KK^*$ &   0.1$\pm$   0.0 &   0.1$\pm$   0.0 &   0.1$\pm$   0.0 & --- \\ \hline
$\pi(1300)\to$ & $\rho\pi$ &  11.2$\pm$  1.9 &  10.1$\pm$  1.8 &   9.4$\pm$  2.1 & seen \\
 & $\rho\omega$ &   0.4$\pm$   0.7 &   0.4$\pm$   0.7 &   0.3$\pm$   0.5 & --- \\
 & $KK^*$ &    0.5$\pm$  2.0 &    0.4$\pm$  1.8 &    0.3$\pm$  1.3 & ---\\ \hline
$\rho(1450)\to$ & $\pi\pi$& 3.8$\pm$   0.3 &  3.4$\pm$   0.2 &  3.4$\pm$   0.2 & seen  \\
 & $KK$ &  0.5$\pm$   0.2 &    0.4$\pm$   0.2 &    0.3$\pm$   0.1 & --- \\
 & $\omega\pi$ & 2.2$\pm$   0.2 &   1.9$\pm$   0.1 &   2.0$\pm$   0.0 & ---\\
 & $\rho\eta$ &   1.4$\pm$   0.2 &   1.3$\pm$   0.1 &   0.8$\pm$   0.1 & ---\\
 & $\rho\eta^\prime$ &    0.1$\pm$   0.0 &    0.1$\pm$   0.0 &    0.0$\pm$   0.0 & --- \\
 & $KK^*$ &  2.8$\pm$   0.9 &  2.5$\pm$   0.8 &  1.4$\pm$   0.5 & ---\\\hline
$\pi_2(1670)\to$ & $(\rho\pi)_{L= 1}$ &   4.6$\pm$   0.2 &   4.1$\pm$   0.2 &   5.5$\pm$   0.2 & 8.6 $\pm$   0.8 $^a$ \\
 & $(\rho\pi)_{L= 3}$ &   7.9$\pm$   0.2 &   7.1$\pm$   0.2 &   5.6$\pm$   0.3 & \\
 & $(\rho\omega)_{S= 1,L= 1}$ &  2.9$\pm$   0.4 &  2.6$\pm$   0.3 &  2.1$\pm$   0.3 & ---\\
 & $(\rho\omega)_{S= 1,L= 3}$ &  1.4$\pm$   0.3 &  1.3$\pm$   0.2 &   0.7$\pm$   0.2 & \\
 & $(KK^*)_{L= 1}$ &   3.5$\pm$   0.4 &   3.1$\pm$   0.4 &   3.0$\pm$   0.5 & 3.2 $\pm$   0.6 $^a$ \\
 & $(KK^*)_{L= 3}$ &   2.1$\pm$   0.5 &   1.9$\pm$   0.5 &   1.0$\pm$   0.3 & \\\hline
%\multicolumn{6}{l}{$^a$ {\footnotesize Experimental values are for the sum over all relevant partial waves.}} \\
\end{tabular}
\end{center}
\end{table}

In tables \ref{tableb} to \ref{tabled}, we have extended the model calculation to examine decays of a few,
semi-randomly selected, radially and orbitally excited states. For these tables, the results correspond to the first
three fits of \ref{tablea}, respectively. We preface our discussion here by reminding
the reader that the assignment of many of the experimentally observed states is still not clear, and it is
not our aim, at least not in this work, to clarify the assignments of any of these states, but simply to see
how well our model for their decays (and implicitly, of their spectrum) works. Thus, we have chosen only a
few of the `better established' states for this discussion. Note that if the predicted value for an amplitude is less than 0.05 MeV$^{1/2}$,
we do not show the amplitude.

In obtaining the numbers for decays with a broad vector meson ($\rho$ or $K^*$; the $\omega$ is less than 10
MeV wide, and we treat it as a narrow state in our calculation) in the final state, we integrate over the
line shape of the broad final state. That is, for such decays, the decay rate is taken as
\begin{equation}
\Gamma_{A\to B V}=\int_0^{k_{max}} dk \frac{k^2\left|M(k)\right|^2\Gamma_V(k)}
{\left(m_A-E_B(k)-E_V(k)\right)^2+\frac{\Gamma_V(k)^2}{4}},
\end{equation}
where $V$ denotes the vector meson, and $\Gamma_V$ is the total width of the vector meson. The errors on the
theoretical numbers in these tables arise from evaluating the amplitudes at the upper and lower limits of the
mass ranges allowed by the experimental error on the masses of the states.

As with the fitted decays, we find that the overall agreement of the model
predictions with the experimental data is quite good. There are a few instances in
which there is noticeable disagreement, and we comment on these below.

Our results for the $a_1(1235)$ are `consistent' with experiment, in that the decay
of this state to $\rho\pi$ is `dominant' (of a total width of about 400 MeV), while
we predict a $\rho\pi$ partial width of about 100 MeV. Not many other exclusive
final states have been identified in the decays of this meson, leaving the
possibility that much of its width comes from decays to excited mesons (or exotics). For the
rest of the decays of the $I=1$ states, the results that we obtain are in reasonable
agreement with what little data there is.

\begin{table} 
\caption{Results from various fits, for states with $I=0$. The key is as in table \protect{\ref{tablea}}.\label{tablec}}
\begin{center}
\begin{tabular}{||rlrrrr||}\hline
\multicolumn{2}{||c}{decay}&\multicolumn{4}{c||}{$\Gamma^{1/2}$ (MeV$^{1/2}$)} \\\cline{3-6}
& & Fit I & Fit II & Fit III & experiment\\ \hline\hline
$f_2(1275)\to$ & $\pi\pi$ &   9.5$\pm$   0.1 &   8.6$\pm$   0.1 &   8.7$\pm$   0.2 &$12.5 \pm 1.3$ \\
& $KK$ &   2.5$\pm$   0.1 &   2.3$\pm$   0.1 &   1.7$\pm$   0.1 & $2.9\pm 0.4$\\
 & $ \eta\eta$ &   0.7$\pm$   0.1 &   0.8$\pm$   0.1 &    0.4$\pm$   0.1 & $0.9 \pm 0.2$ \\\hline
$f_1(1280)\to$ & $KK^*$ &   0.5$\pm$   0.1 &   0.4$\pm$   0.1 &   0.4$\pm$   0.1 &  not seen \\\hline
$f_0(1300)\to$ & $\pi\pi$ &  11.6$\pm$  5.7 &  10.5$\pm$  5.1 &  15.6$\pm$  6.6 & $16.0\pm 6.7$\\
& $KK$ &   9.5$\pm$  3.6 &   8.6$\pm$  3.2 &  10.3$\pm$  2.5 & $4.5 \pm 2.1$\\
 & $ \eta\eta$ &  6.1$\pm$ 6.1 &  5.5$\pm$ 5.5 &  6.2$\pm$ 6.2 & seen \\ \hline
$f_2(1525)\to$ & $KK$ &   9.4$\pm$   0.2 &   8.5$\pm$   0.2 &   8.4$\pm$   0.2 & $7.3\pm 1.6$ \\
 & $ \eta\eta$ &   2.6$\pm$   0.1 &   2.4$\pm$   0.1 &   2.3$\pm$   0.1 & $4.6 \pm 1.1$\\
 & $ \eta\eta^\prime$ &    0.0$\pm$   0.1 &    0.0$\pm$   0.1 &    0.0$\pm$   0.0 & ---\\
 & $ KK^*$ &  3.2$\pm$   0.5 &  2.9$\pm$   0.5 &  2.2$\pm$   0.4 & --- \\ \hline
$f_1(1510)\to$ & $ (KK^*)_{L= 0}$ &  9.4$\pm$   0.5 &  8.5$\pm$   0.4 &  8.5$\pm$   0.5 & seen 
\tablenote{Experimental values are for the sum over all relevant partial waves.} \\
& $ (KK^*)_{L= 2}$ &   1.9$\pm$   0.3 &   1.7$\pm$   0.3 &   1.3$\pm$   0.2 & \\\hline
$f_0(1500)\to$ & $KK$ &  12.6$\pm$   0.3 &  11.4$\pm$   0.3 &  15.5$\pm$   0.3 & --- \\
 & $ \eta\eta$ &  7.3$\pm$   0.1 &  6.6$\pm$   0.1 &  8.5$\pm$   0.1 & seen \\
 & $ \eta\eta^\prime$ &    0.0$\pm$  3.9 &    0.0$\pm$  3.6 &    0.0$\pm$  3.6 & seen \\\hline
$\phi(1680)\to$ & $KK$ &   5.6$\pm$   0.2 &   5.0$\pm$   0.2 &   6.8$\pm$   0.1 & seen\\
 & $ \eta\eta$ &   1.5$\pm$   0.1 &   1.4$\pm$   0.1 &   1.8$\pm$   0.0 & --- \\
 & $ \eta\eta^\prime$ &   1.4$\pm$  1.4 &   1.2$\pm$  1.2 &   0.8$\pm$  0.8 & --- \\
 & $ KK^*$ & 12.6$\pm$  1.9 & 11.4$\pm$  1.7 & 11.2$\pm$  2.0 &  dominant \\\hline
%\multicolumn{6}{l}{$^a$ {\footnotesize Experimental values are for the sum over all relevant partial waves.}} \\
\end{tabular}
\end{center}
\end{table}

In the $I=0$ channels, our predictions are in quite good agreement with experiment, except for the decays of
the $f_0(1300)$. However, the assignment of this state is so uncertain, (e. g.; there still is the question of
whether this is a single state or not) that interpretation of the discrepancy is difficult.
In this sector there is one decay that is worthy of some comment. At its nominal mass, the $f_0(1500)$ will
not decay to $\eta\eta^\prime$, yet the partial width for this decay has been listed as `large (dominant)' \cite{pdg1}.
However, we see that if we change the mass of this state by 10 MeV in our model (a small amount, considering 
how broad $0^{++}$ states are expected to be), the amplitude for decay to $\eta\eta^\prime$ grows rapidly. 
This is not surprising, as this is an $S$-wave amplitude, which grows roughly as $(M-M_{\rm
th})^{1/2}$, where $M$ is the mass of parent and $M_{\rm th}$ is the threshold for the decay. This would
suggest that either (a) this state is more massive than reported in \cite{pdg}; or (b) the amplitude grows so rapidly
that we should integrate over the line-shape of at least the $\eta^\prime$ in calculating the amplitude. This
decay also illustrates some of the importance of threshold effects.

\begin{table}
\caption{Results from various fits, for kaons. The key is as in table \protect{\ref{tablea}}.\label{tabled}}
\begin{center}
\begin{tabular}{||rlrrrr||}\hline
\multicolumn{2}{||c}{decay}&\multicolumn{4}{c||}{$\Gamma^{1/2}$ (MeV$^{1/2}$)} \\\cline{3-6}
& & Fit I & Fit II & Fit III & experiment \\ \hline\hline 
$K_1(^1P_1)\to$ & $K\omega$ &    0.0$\pm$  2.9 &    0.0$\pm$  2.6 &    0.0$\pm$  2.2 & $3.1\pm 0.9$ 
\tablenote{The experimental numbers are for the fully mixed states, while the 
theoretical numbers for the $^1P_1$ and $^3P_1$ unmixed states.} 
\tablenote{Experimental values are for the sum over all relevant partial waves.}\\
& $(K\rho)_{L= 0}$ &   3.0$\pm$   0.3 &   2.7$\pm$   0.3 &   2.5$\pm$   0.3 & 6.1 $\pm$  1.1 $^{a\,\,b}$\\
& $(K\rho)_{L= 2}$ &    0.9$\pm$   0.1 &    0.8$\pm$   0.1 &    0.6$\pm$   0.1 & \\
& $(\pi K^*)_{L= 0}$ &   4.7$\pm$   0.1 &   4.2$\pm$   0.1 &   4.2$\pm$   0.1 & 3.8 $\pm$  1.0 $^{a\,\,b}$\\
& $(\pi K^*)_{L= 2}$ &   1.8$\pm$   0.1 &   1.6$\pm$   0.1 &   1.1$\pm$   0.1 & \\\hline
$K_1(^3P_1)\to$ & $(K\omega)_{L= 0}$ &   6.9$\pm$   0.1 &   6.2$\pm$   0.1 &   6.1$\pm$   0.0 & $1.3 \pm 0.8$ $^{a\,\,b}$\\
& $(K\omega)_{L= 2}$ &  1.3$\pm$   0.1 &  1.2$\pm$   0.1 &   0.7$\pm$   0.1 & \\
& $(K\rho)_{L= 0}$ &  8.5$\pm$   0.2 &  7.7$\pm$   0.2 &  7.7$\pm$   0.3 & 2.3 $\pm$  1.2 $^{a\,\,b}$\\
& $(K\rho)_{L= 2}$ &   2.4$\pm$   0.2 &   2.2$\pm$   0.2 &   1.6$\pm$   0.1 & \\
& $(\pi K^*)_{L= 0}$ & 10.3$\pm$   0.1 &  9.3$\pm$   0.1 & 10.5$\pm$   0.2 & 12.8 $\pm$   0.9 $^{a\,\,b}$\\
& $(\pi K^*)_{L= 2}$ &   4.5$\pm$   0.2 &   4.1$\pm$   0.2 &   3.1$\pm$   0.2 & \\\hline
$K^*(1410)\to$ & $K\pi$ &  1.3$\pm$   0.1 &  1.2$\pm$   0.1 &  1.1$\pm$   0.0 & $3.9\pm 0.8$ $^b$ \\
& $K\eta$ &    0.1$\pm$   0.0 &    0.1$\pm$   0.0 &    0.8$\pm$   0.0 & --- \\
& $K\omega$ &   2.8$\pm$   0.1 &   2.5$\pm$   0.1 &   1.5$\pm$   0.1 & --- \\
& $K\rho$ &  3.8$\pm$   0.1 &  3.4$\pm$   0.1 &  2.3$\pm$   0.1 & $<$ 4.0 $^b$\\
& $\pi K^*$ &  4.9$\pm$   0.1 &  4.4$\pm$   0.0 &  3.3$\pm$   0.1 &$>$ 9.5 $^b$ \\\hline
$K_0(1430)\to$ & $K\pi$ &   7.5$\pm$   0.2 &   6.8$\pm$   0.2 &  10.4$\pm$   0.3 & $16.3 \pm 2.0$ \\
& $K\eta$ & 11.2$\pm$   0.1 & 10.2$\pm$   0.1 & 13.2$\pm$   0.1 & ---\\\hline
$K_2(1425)\to$ & $K\pi$ &   7.9$\pm$   0.1 &   7.1$\pm$   0.1 &   7.1$\pm$   0.2 & $7.2\pm 1.5$ \\
& $K\eta$ &  4.2$\pm$   0.1 &  3.7$\pm$   0.1 &  3.1$\pm$   0.1 & 0.4 $\pm$   0.4\\
& $K\omega$ &   2.1$\pm$   0.2 &   1.9$\pm$   0.2 &   1.3$\pm$   0.1 & $1.7 \pm 0.6$ \\
& $K\rho$ &  3.7$\pm$   0.2 &  3.3$\pm$   0.2 &  2.5$\pm$   0.2 & 3.1 $\pm$   0.2\\
& $\pi K^*$ &  6.4$\pm$   0.1 &  5.8$\pm$   0.1 &  4.4$\pm$   0.1 & 5.2 $\pm$   0.3\\ \hline
$K(1460)\to$ & $ K\omega$ &  5.8$\pm$   0.7 &  5.2$\pm$   0.6 &  4.1$\pm$   0.6 & --- \\
& $K\rho$ &   8.7$\pm$   0.8 &   7.8$\pm$   0.7 &   6.6$\pm$   0.8 & $\simeq$ 5.8\\
& $\pi K^*$ &  12.2$\pm$   0.9 &  11.0$\pm$   0.8 &  10.1$\pm$   0.5 & $\simeq$ 10.1\\\hline
$K_2(1580)\to$ & $(K\omega)_{L= 1}$ &  2.4$\pm$   0.2 &  2.2$\pm$   0.2 &  2.1$\pm$   0.5 & --- \\
& $(K\omega)_{L= 3}$ &  1.5$\pm$   0.9 &  1.3$\pm$   0.8 &   0.6$\pm$   0.7 & \\
& $(K\rho)_{L= 1}$ &   4.4$\pm$  1.5 &   4.0$\pm$  1.3 &   4.0$\pm$  1.6 & --- \\
& $(K\rho)_{L= 3}$ &   3.1$\pm$  1.7 &   2.8$\pm$  1.5 &   1.6$\pm$  1.0 & \\
& $(\pi K^*)_{L= 1}$ &   3.0$\pm$   0.5 &   2.7$\pm$   0.4 &   2.8$\pm$   0.9 & seen $^b$ \\
& $(\pi K^*)_{L= 3}$ &   2.3$\pm$  2.0 &   2.1$\pm$  1.8 &   1.2$\pm$  1.5 & \\
& $(\omega K^*)_{S= 1,L= 1}$ &    0.0$\pm$   0.8 &    0.0$\pm$   0.7 &    0.0$\pm$   0.4 & --- \\
& $(\omega K^*)_{S= 1,L= 3}$ &    0.0$\pm$   0.1 &    0.0$\pm$   0.1 &    0.0$\pm$   0.0 &  \\\hline
$K_1(1680)\to$ & $K\pi$ &   3.6$\pm$   0.1 &   3.2$\pm$   0.1 &   5.2$\pm$   0.1 & $11.2 \pm 3.0$\\
& $K\eta$ &  4.0$\pm$   0.1 &  3.7$\pm$   0.0 &  5.0$\pm$   0.0 & ---\\
& $K\omega$ &  2.0$\pm$   0.1 &  1.8$\pm$   0.1 &  2.1$\pm$   0.0 & --- \\
& $K\rho$ &   3.4$\pm$   0.1 &   3.1$\pm$   0.1 &   3.6$\pm$   0.0 & 10.1 $\pm$  2.5\\
& $\pi K^*$ &   3.2$\pm$   0.2 &   2.9$\pm$   0.1 &   3.8$\pm$   0.3 & 9.8 $\pm$   0.9\\ 
& $(\omega K^*)_{S= 0,L= 1}$ &   1.2$\pm$   0.6 &   1.1$\pm$   0.5 &    0.7$\pm$   0.3 & ---\\
& $(\omega K^*)_{S= 2,L= 1}$ &   0.5$\pm$   0.3 &   0.5$\pm$   0.2 &   0.3$\pm$   0.1 & \\
& $(\omega K^*)_{S= 2,L= 3}$ &   0.4$\pm$   0.2 &   0.4$\pm$   0.2 &   0.1$\pm$   0.1 & \\\hline
$K_2(1770)\to$ & $(K\omega)_{L= 1}$ &   2.3$\pm$   0.1 &   2.1$\pm$   0.1 &   2.8$\pm$   0.0 & seen $^b$ \\
& $(K\omega)_{L= 3}$ &  2.6$\pm$   0.1 &  2.4$\pm$   0.1 &  1.7$\pm$   0.1 & \\
& $(K\rho)_{L= 1}$ &  4.3$\pm$   0.0 &  3.9$\pm$   0.0 &  4.9$\pm$   0.0 & --- \\
& $(K\rho)_{L= 3}$ &   5.1$\pm$   0.3 &   4.6$\pm$   0.3 &   3.6$\pm$   0.3 &  \\
& $(\pi K^*)_{L= 1}$ &  4.9$\pm$   0.1 &  4.5$\pm$   0.1 &  6.3$\pm$   0.2 & --- \\
& $(\pi K^*)_{L= 3}$ &   6.8$\pm$   0.5 &   6.2$\pm$   0.4 &   5.0$\pm$   0.4 & \\ 
& $(\omega K^*)_{S= 2,L= 1}$ &   3.5$\pm$   0.7 &   3.2$\pm$   0.6 &   2.2$\pm$   0.4 & --- \\
& $(\omega K^*)_{S= 2,L= 3}$ &   1.5$\pm$   0.4 &   1.4$\pm$   0.3 &    0.5$\pm$   0.1 & --- \\\hline
$K_3(1770)\to$ & $K\pi$ &   4.8$\pm$   0.1 &   4.3$\pm$   0.1 &   4.3$\pm$   0.1 & $5.6\pm 0.8$ \\
& $K\eta$ &  2.9$\pm$   0.1 &  2.7$\pm$   0.0 &  2.1$\pm$   0.1 & 3.6 $\pm$   0.8\\
& $K\omega$ &   2.9$\pm$   0.1 &   2.6$\pm$   0.1 &   1.9$\pm$   0.1 & $3.6 \pm 0.5$\\
& $K\rho$ &  6.0$\pm$   0.3 &  5.4$\pm$   0.3 &  4.3$\pm$   0.3 & 8.6 $\pm$   0.8\\
& $\pi K^*$ &  8.0$\pm$   0.5 &  7.2$\pm$   0.5 &  5.8$\pm$   0.5 & 6.7 $\pm$   0.7\\ 
& $(\omega K^*)_{S= 0,L= 3}$ &    0.4$\pm$   0.1 &    0.4$\pm$   0.1 &    0.2$\pm$   0.0 & --- \\
& $(\omega K^*)_{S= 2,L= 1}$ &  6.7$\pm$  1.5 &  6.1$\pm$  1.4 &  4.3$\pm$  1.0 &  \\
& $(\omega K^*)_{S= 2,L= 3}$ &  1.0$\pm$   0.3 &   0.9$\pm$   0.2 &   0.3$\pm$   0.1 &\\\hline
%\multicolumn{6}{l}{$^a$ {\footnotesize The experimental numbers are for the fully mixed states, while the 
%theoretical numbers}}\\
%\multicolumn{6}{l}{{\footnotesize are for the $^1P_1$ and $^3P_1$ unmixed states.}}\\
%\multicolumn{6}{l}{$^a$ {\footnotesize The experimental numbers are for the fully mixed states, while the 
%theoretical numbers for the $^1P_1$ and $^3P_1$ unmixed states.}}\\
%\multicolumn{6}{l}{$^b$ {\footnotesize Experimental values are for the sum over all relevant partial waves.}}
\end{tabular}
\end{center}
%\end{longtable}
\end{table}

In the kaon sector, our predictions are again in reasonable agreement with experiment. In the case of the
$K_1$'s, we can not really compare theory and experiment, since we have not included mixing. This can easily
be taken into account in the decays to the narrow states, but not for the broad states, since an integral over
the partial {\em width} is required to evaluate the amplitude. Thus, simply constructing a linear combination
of the reported amplitudes will be ignoring potentially large interference effects.

The place where our model `fails' most noticeably is in describing the decays of the $K_1(1680)$. There, our
amplitudes are consistently smaller than the experimental ones. Note, however, that for the three measured
amplitudes, our results agree `qualitatively' with the experiments, in that the 3 amplitudes are all predicted
to be of roughly the same size.

One other aspect of our results requires some comment. Our results have been largely independent of the size
of the expansion basis used, as well as of the choice of potential used to generate the wave functions. This
may be understood by examining what part of the wave function provides the dominant contribution to any one of
the amplitudes. It is apparent that these decays are neither fully `long-distance' nor `short-distance'
processes, but more like `intermediate-distance' (1-5 GeV$^{-1}$) processes. Changes in the basis size have their main 
effect on the shape of the wave function in the region of the origin, as well as on the long distance tail. 
By changing the potential from Al1 to AP1, we change the same regions of the wave function.

\section{Conclusion}

We have examined the predictions of a quark model for the strong decays of several mesons.
While this study was not exhaustive, it has been enough to show that the model enjoys some
success, but that some improvements may be possible. The main improvements that can be made
exist at the level of the meson wave functions. 

At present, there is no spin-orbit term in
either of the potentials used, nor is there a tensor interaction. Both of these will lead to
modifications of the wave functions that will, in turn, lead to changes in the strong decay
amplitudes predicted by the model. These terms will also induce various mixings among the
states, which will also modify the wave functions. Such a study is underway, and the use of
the modified wave functions in a strong decay analysis is left as a possible future endeavor.
It would certainly be of interest to apply such wave functions to a more exhaustive examination
of the strong decays of mesons, as there are several puzzles outstanding in the meson sector,
particularly for mesons with masses between 1.0 and 2.0 GeV. To date, models of this kind
have been our main source of insight into the meson spectrum. Barring a radical revolution in
the near future, they will remain so, at least for some time to come.

\section*{Acknowledgement}

W. R. acknowledges the hospitality of Institut des Sciences Nucl\'eaires,
Grenoble, france, where most of this work was done, as well as the support of
DOE under contracts DE-AC05-84ER40150 and DE
FG05-94ER40832, and the NSF under award PHY 9457892.

\end{document}